\documentclass[%
    prd,      
    preprint, 
    showkeys, 
    superscriptaddress, 
    nofootinbib 
]{revtex4}

\usepackage{amsmath,amsfonts,amssymb}
\usepackage{booktabs} 
\usepackage{csquotes} 
\usepackage{enumitem}
\usepackage{graphicx}
\usepackage{tabularx}
\usepackage{multirow}
\graphicspath{{./figures/}}

\usepackage{color}
\definecolor{verdes}{cmyk}{0.92,0,0.59,0.4}  
\definecolor{verdec}{cmyk}{0.92,0,0.59,0.15} 

\usepackage{hyperref}
\hypersetup{
    colorlinks=true,
    linkcolor=verdec,
    urlcolor=verdec,
    citecolor=verdec
}

\newcommand{\gev}{{\;{\rm GeV}}}
\newcommand{\tev}{{\;{\rm TeV}}}

\newcommand{\beq}{\begin{equation}}
\newcommand{\eeq}{\end{equation}}
\newcommand{\bea}{\begin{eqnarray}}
\newcommand{\eea}{\end{eqnarray}}
\newcommand{\barr}{\begin{array}}
\newcommand{\earr}{\end{array}}
\newcommand{\bc}{\begin{center}}
\newcommand{\ec}{\end{center}}
\newcommand{\bit}{\begin{itemize}}
\newcommand{\eit}{\end{itemize}}
\newcommand{\ben}{\begin{enumerate}}
\newcommand{\een}{\end{enumerate}}

\newcommand{\package}[1]{\textsc{\small #1}}

\newcommand{\mh}{m_{h}}

\newcommand{\mch}{M_{H^\pm}}
\newcommand{\mhh}{M_{H}}
\newcommand{\ma}{M_{A}}

\newcommand{\tb}{t_\beta}

\newcommand{\cba}{c_{\beta-\alpha}}

\begin{document}
\preprint{KIAS-P26045}

\title{Cutoff Scales in the Type-I 2HDM with Strongly First-Order Electroweak Phase Transitions: One-Step versus Multistep}

\author{Jin-Hwan Cho}
\email{chof@nims.re.kr}
\affiliation{National Institute for Mathematical Sciences, Daejeon 34047, Republic of Korea} 

\author{Dongjoo Kim}
\email{dongjookim.phys@gmail.com}
\affiliation{National Institute for Mathematical Sciences, Daejeon 34047, Republic of Korea} 
\affiliation{Department of Physics, Konkuk University, Seoul 05029, Republic of Korea} 

\author{Jinheung Kim}
\email{jhkim1216@kias.re.kr}
\affiliation{School of Physics, Korea Institute for Advanced Study,  Seoul 02455, Republic of Korea} 

\author{Jeonghyeon Song}
\email{jhsong@konkuk.ac.kr}
\affiliation{Department of Physics, Konkuk University, Seoul 05029, Republic of Korea} 

\begin{abstract}
We investigate the UV viability of strongly first-order electroweak phase transitions (SFOEWPTs) in the Type-I two-Higgs-doublet model (2HDM), considering both the Normal and Inverted Scenarios (NS and IS). For each scenario, we scan $5\times10^6$ physical parameter points and determine the cutoff scale $\Lambda_{\rm c}$ through a two-loop renormalization-group analysis, where $\Lambda_{\rm c}$ is set by the first violation of perturbativity, tree-level unitarity, or vacuum stability. For one-step transitions, the SFOEWPT strength and high-scale UV validity exhibit a pronounced tension: the maximal transition strength $\xi_p$ decreases with increasing $\Lambda_{\rm c}$. Requiring $\Lambda_{\rm c}>10~\text{TeV}$ limits the transition strength to $\xi_p\lesssim2.7$ in the NS and $\xi_p\lesssim1.8$ in the IS, while requiring $\xi_p>1$ restricts the cutoff scale to $\Lambda_{\rm c}\lesssim \mathcal{O}(10^6)~\text{GeV}$ in both scenarios. Multistep transitions exhibit qualitatively different behavior. Two-step SFOEWPTs, found in appreciable numbers only in the IS, can remain theoretically consistent up to $\Lambda_{\rm c}\sim \mathcal{O}(10^{15})~\text{GeV}$ while reaching $\xi_p\simeq7$, without exhibiting the pronounced $\xi_p$--$\Lambda_{\rm c}$ anticorrelation characteristic of one-step transitions. Imposing UV validity also sharpens the phenomenologically viable parameter space. In particular, increasing the minimum allowed cutoff scale for two-step SFOEWPTs in the IS favors a light scalar spectrum and moderate $\tan\beta$, providing a promising target for current and future collider searches.
\end{abstract}


\maketitle

\tableofcontents


\section{Introduction}

A strongly first-order electroweak phase transition
(SFOEWPT)~\cite{Huet:1994jb,Kajantie:1996mn} has attracted considerable
interest, as it can provide the departure from thermal
equilibrium~\cite{Sakharov:1967dj} required for generating the observed
baryon asymmetry of the Universe, characterized by the baryon-to-photon
ratio $n_B/n_\gamma\approx6\times10^{-10}$~\cite{Planck:2015fie}.
An SFOEWPT would also source a stochastic gravitational-wave
background~\cite{Weir:2017wfa,Caprini:2018mtu}, potentially observable
at future space-based interferometers and thereby providing a
complementary probe of the thermal
history in the early Universe~\cite{LISA:2017pwj,Cutting:2018tjt,Guo:2020grp,Schmitz:2020syl,Caprini:2015zlo,Caprini:2019egz}.
Since the EWPT in the Standard Model (SM) proceeds via a smooth
crossover~\cite{Huet:1994jb,Kajantie:1996mn,Csikor:1998eu}, realizing
an SFOEWPT motivates an extended Higgs sector beyond the SM
(BSM)~\cite{Bochkarev:1990fx,Dorsch:2013wja,Basler:2016obg,Fuyuto:2017ewj,Bernon:2017jgv,Kainulainen:2019kyp,Kanemura:2022ozv,Bittar:2025lcr}.

An especially interesting possibility in extended Higgs sectors is a
multistep phase transition, in which the Universe passes through one or
more intermediate metastable vacua before reaching the electroweak
minimum~\cite{Land:1992sm,Patel:2012pi,Inoue:2015pza,Blinov:2015sna,Bian:2017wfv,Chao:2017vrq,Ramsey-Musolf:2017tgh,Chala:2018opy,Morais:2019fnm,Fabian:2020hny,Aoki:2021oez,Cao:2022ocg,Goncalves:2021egx,Karan:2023kyj,Si:2024vrq,Coutinho:2024zyp,Bhatnagar:2025jhh,Conaci:2026djb}.
Multistep transitions are not merely rarer variants of the one-step
case; their properties can differ qualitatively.
In particular, correlations established for one-step transitions may
cease to hold once intermediate vacua are involved.
The zero-temperature correlation between the vacuum-uplifting
quantity $\Delta F_0$~\cite{Dorsch:2017nza,Su:2020pjw,Goncalves:2021egx}
and the transition strength, for instance, holds for one-step transitions
but is lost in the multistep case~\cite{Lee:2025hgb}.
Whether this qualitative distinction extends to other correlations
remains an open question.

A closely related but largely unexplored issue concerns the cutoff scale
$\Lambda_{\rm c}$ of SFOEWPT parameter points, defined as the highest
energy scale up to which the underlying BSM theory remains theoretically
consistent.
As the couplings evolve under their renormalization-group equations (RGEs), a
parameter point satisfying perturbativity, unitarity, and vacuum stability
at the electroweak scale may violate one of these conditions at
$\Lambda_{\rm c}$.
Beyond this scale, the 2HDM can no longer be regarded as a theoretically
consistent description without additional UV physics. 
Regarding the UV viability of SFOEWPTs, several important questions remain
unanswered to our knowledge:
\textit{how is the cutoff scale related to the EWPT strength, how does this
relation depend on whether the thermal history is one-step or multistep,
and how do the requirements $\Lambda_{\rm c}>\textit{1}~ TeV$ and $\textit{10}~ TeV$ restrict
the viable SFOEWPT parameter space, including the BSM Higgs masses and
the mixing angles?}

As a concrete BSM framework, we consider the Type-I
two-Higgs-doublet model (2HDM)~\cite{Branco:2011iw}.
The 2HDM is one of the simplest extensions of the SM and contains five
physical scalar states: two CP-even scalars ($h,H$), one CP-odd scalar
($A$), and a pair of charged Higgs bosons ($H^\pm$).
A softly broken discrete $Z_2$ symmetry is imposed to avoid tree-level
flavor-changing neutral currents, and we focus on the Type-I Yukawa
structure, which remains comparatively weakly constrained, particularly
by radiative $B$-meson
decays~\cite{Haller:2018nnx,Misiak:2020vlo,Biekotter:2024ykp,
Biekotter:2025fjx}.
The observed $125~\text{GeV}$ scalar
$h_{125}$~\cite{ATLAS:2012yve,CMS:2012qbp} may be identified with either
the lighter CP-even state, defining the Normal Scenario (NS), or the
heavier CP-even state, defining the Inverted Scenario (IS).

Within this framework, we perform a global analysis of the UV viability of
one-step and multistep SFOEWPTs in both the NS and IS.\footnote{Previous work on the UV viability of
SFOEWPTs has been limited in scope. Ref.~\cite{Biekotter:2022kgf}
examined the perturbativity cutoff $\Lambda_{4\pi}$ of SFOEWPT regions,
where $\Lambda_{4\pi}$ is the energy scale at which one of the
quartic couplings exceeds the perturbative bound $4\pi$, but only for
the NS Type-II 2HDM in the exact Higgs alignment limit.}
For each scenario, we scan $5\times10^6$ physical parameter points,
identify SFOEWPT points---those for which the percolation order parameter
satisfies $\xi_p>1$---with one-, two-, and three-step thermal histories,
and determine $\Lambda_{\rm c}$ through a two-loop RGE analysis.
Compared with our previous studies of the NS~\cite{Cho:2026pfw} and
IS~\cite{Lee:2025hgb}, the present analysis enlarges the scan samples,
adopts a common setup for both scenarios, and incorporates the RGE
evolution needed to assess their UV validity.

We find a sharp distinction between one-step and multistep SFOEWPTs.
For one-step transitions, which overwhelmingly dominate the SFOEWPT
samples, the maximal $\xi_p$ decreases with increasing $\Lambda_{\rm c}$:
requiring a higher cutoff lowers the upper bound on $\xi_p$, whereas
requiring $\xi_p>1$ restricts $\Lambda_{\rm c}$ to at most
$\mathcal{O}(10^6)\gev$ in both the NS and IS.
By contrast, two-step SFOEWPTs, found in appreciable numbers only in the
IS, exhibit qualitatively different behavior.
The pronounced $\xi_p$--$\Lambda_{\rm c}$ anticorrelation observed for
one-step transitions is absent: cutoff scales can reach
$\mathcal{O}(10^{15})\gev$, while the two-step sample also contains
transitions as strong as $\xi_p\simeq7$.
Within this two-step region, imposing a higher minimum cutoff scale
further reshapes the surviving parameter space, favoring lighter BSM
Higgs bosons and moderate $\tan\beta$.
The resulting high-cutoff region therefore provides a promising target
for current and future collider searches.

\section{Review of SFOEWPT in the 2HDM}
\label{sec-review}

The 2HDM extends the SM scalar sector by introducing two complex
$SU(2)_L$ doublets, $\Phi_1$ and $\Phi_2$, each with hypercharge
$Y=1/2$~\cite{Branco:2011iw}:
\begin{equation}
\label{eq:phi:fields}
\Phi_i =
\begin{pmatrix}
w_i^+ \\
\dfrac{v_i+\rho_i+i\eta_i}{\sqrt{2}}
\end{pmatrix},
\qquad (i=1,2),
\end{equation}
where $v_1$ and $v_2$ are the vacuum expectation values (VEVs) of
$\Phi_1$ and $\Phi_2$, respectively. The combined VEV,
$v=\sqrt{v_1^2+v_2^2}\approx246~\text{GeV}$, drives electroweak symmetry
breaking, while their ratio defines the parameter
$t_\beta=v_2/v_1$. For notational simplicity, we use
$s_x=\sin x$, $c_x=\cos x$, and $t_x=\tan x$.

A discrete $Z_2$ symmetry, under which $\Phi_1\to\Phi_1$ and
$\Phi_2\to-\Phi_2$, is imposed to forbid tree-level flavor-changing
neutral currents~\cite{Glashow:1976nt,Paschos:1976ay}.
The CP-conserving scalar potential with soft $Z_2$ breaking is
\begin{equation}
\label{eq:VPhi}
\begin{split}
V_\Phi(\Phi_1,\Phi_2)
&=m^2_{11}\Phi^\dagger_1\Phi_1
 +m^2_{22}\Phi^\dagger_2\Phi_2
 -m^2_{12}(\Phi^\dagger_1\Phi_2+{\rm H.c.}) \\[3pt]
&\quad
 +\tfrac{1}{2}\lambda_1(\Phi^\dagger_1\Phi_1)^2
 +\tfrac{1}{2}\lambda_2(\Phi^\dagger_2\Phi_2)^2
 +\lambda_3(\Phi^\dagger_1\Phi_1)(\Phi^\dagger_2\Phi_2) \\[3pt]
&\quad
 +\lambda_4(\Phi^\dagger_1\Phi_2)(\Phi^\dagger_2\Phi_1)
 +\tfrac{1}{2}\lambda_5
 \bigl[(\Phi^\dagger_1\Phi_2)^2+{\rm H.c.}\bigr],
\end{split}
\end{equation}
where $m_{12}^2$ softly breaks the $Z_2$ symmetry.
The physical scalar spectrum consists of two CP-even states ($h,H$),
one CP-odd state ($A$), and a charged pair ($H^\pm$), related to the
interaction eigenstates in \autoref{eq:phi:fields} through the mixing
angles $\alpha$ and $\beta$~\cite{Lee:2025hgb}.
The SM Higgs direction is given by
$h_{\rm SM}=s_{\beta-\alpha}h+\cba H$.
Accordingly, the observed $125~\text{GeV}$ scalar
$h_{125}$~\cite{ATLAS:2012yve,CMS:2012qbp} may be identified either
with the lighter CP-even state $h$, defining the Normal Scenario (NS),
or with the heavier state $H$, defining the Inverted Scenario (IS).
Phenomenologically viable points therefore lie close to the Higgs
alignment limit, with $|s_{\beta-\alpha}|\simeq1$ in the NS and
$|s_{\beta-\alpha}|\simeq0$ in the IS.

The dynamics of the EWPT in the early Universe is governed by the
finite-temperature effective potential $V_{\rm eff}(\vec{w},T)$, where
$\vec{w}=(w_1,w_2,w_3)$ represents the spatially homogeneous classical
background fields~\cite{Espinosa:1992kf,Herring:2024pqa,Chakrabortty:2024wto,Masina:2025pnp}:
\begin{equation}
\label{eq-classical-field}
\Phi_1^c=\frac{1}{\sqrt{2}}
\begin{pmatrix}
0\\ w_1
\end{pmatrix},
\qquad
\Phi_2^c=\frac{1}{\sqrt{2}}
\begin{pmatrix}
0\\ w_2+i w_3
\end{pmatrix}.
\end{equation}
At one-loop level,
\begin{equation}
\label{eq-Veff}
V_{\rm eff}(\vec{w},T)
=
V_{\rm tree}(\vec{w})
+V_{\rm CW}(\vec{w})
+V_{\rm CT}(\vec{w})
+V_T(\vec{w},T),
\end{equation}
where $V_{\rm tree}$, $V_{\rm CW}$, $V_{\rm CT}$, and $V_T$ denote the
tree-level, Coleman--Weinberg, counterterm, and thermal contributions,
respectively. Their explicit forms are given in our previous
work~\cite{Lee:2025hgb}.

An important aspect of evaluating $V_{\rm eff}(\vec{w},T)$ is the
treatment of thermal daisy effects. At finite temperature, bosonic
Matsubara zero modes lead to infrared divergences in the perturbative
expansion, causing the naive fixed-order loop expansion to break down
near the phase transition~\cite{Dolan:1973qd,Linde:1980ts,Parwani:1991gq,Arnold:1992rz}.
Thermal resummation is therefore required to reorganize the perturbative
expansion by incorporating thermal screening effects.
Between the conventional Parwani~\cite{Parwani:1991gq} and
Arnold--Espinosa~\cite{Arnold:1992rz} daisy-resummation schemes, we
adopt the Parwani method, motivated by its better agreement with results
obtained using advanced partial-dressing
formalisms~\cite{Boyd:1993tz,Curtin:2016urg,Curtin:2022ovx,Bahl:2024ykv,Bittar:2025lcr} 
and a two-particle-irreducible Hartree
effective potential~\cite{Banik:2026ljf}.
In the Parwani scheme, thermal screening effects are incorporated by
replacing the field-dependent tree-level mass eigenvalues entering the
one-loop effective potential with thermally corrected ones,
\begin{equation}
M_i^2(\vec{w},T)=m_i^2(\vec{w})+\Pi_i(T),
\end{equation}
where $\Pi_i(T)$ denotes the thermal self-energy of the corresponding
bosonic mode.

The finite-temperature effective potential determines several
characteristic temperatures during the vacuum evolution.
At the critical temperature $T_c$, the electroweak-symmetric and broken
minima become degenerate and are separated by a potential barrier.
At the nucleation temperature $T_n$, thermal tunneling becomes
sufficiently efficient that the number of critical bubbles nucleated
within a Hubble volume is of order unity~\cite{Coleman:1977py,Callan:1977pt,Linde:1981zj,Bardsley:2021lmq,Lofgren:2021ogg,Athron:2022mmm}.
The bubbles subsequently expand and eventually percolate, defining the
percolation temperature $T_p$.

Electroweak baryogenesis requires the EWPT to be strongly first order so
that, after the transition, the electroweak symmetry-breaking VEV remains
sufficiently large compared with the temperature.
Since the sphaleron energy in the broken phase scales with this
temperature-dependent VEV, a sufficiently large ratio $v(T)/T$ suppresses
baryon-number-violating sphaleron processes behind the expanding bubble
wall and prevents the generated baryon asymmetry from being washed
out~\cite{Quiros:1999jp}.

The strength of the EWPT is commonly characterized by the order
parameter $\xi$, defined as the ratio of the symmetry-breaking VEV to
the temperature.
Evaluating this quantity at the critical, nucleation, and percolation
temperatures defines $\xi_c$, $\xi_n$, and $\xi_p$, respectively.
Both $\xi_c$ and $\xi_n$, however, have limitations as measures of the
physically realized transition.
The static quantity $\xi_c$ does not determine whether the transition is
dynamically realized and may therefore include vacuum-trapped
configurations that do not complete the transition~\cite{Biekotter:2022kgf}.
Although $\xi_n$ incorporates tunneling dynamics, $T_n$
marks only the onset of efficient bubble nucleation rather than the
macroscopic progress of the transition.
By contrast, $T_p$ corresponds to a stage at which the broken phase
occupies a macroscopic fraction of the Universe.
We therefore use the percolation order parameter $\xi_p$ to identify an
SFOEWPT.

A distinctive possibility in extended Higgs sectors is a multistep phase
transition~\cite{Land:1992sm,Patel:2012pi,Inoue:2015pza,Blinov:2015sna,Ramsey-Musolf:2017tgh,Chala:2018opy,Morais:2019fnm,Fabian:2020hny}, in which the Universe passes through one or more
intermediate metastable vacua before reaching the final electroweak
vacuum.
For each transition stage, the dynamical order parameter at percolation
is defined as
\begin{equation}
\xi_p\equiv\frac{\Delta v(T_p)}{T_p},
\end{equation}
where $\Delta v(T_p)$ denotes the magnitude of the change in the VEVs
between the two successive phases.
We identify an SFOEWPT by the condition
\beq
\label{eq-SFOEWPT-condition}
\text{SFOEWPT:}\quad \xi_p>1.
\eeq

Another important characteristic of an FOEWPT is cosmic supercooling,
which occurs when the Universe remains in a metastable phase as it cools
below $T_c$ before the transition proceeds through bubble nucleation and
growth~\cite{Kobakhidze:2017mru,Ellis:2018mja,Athron:2022mmm}.
Greater supercooling generally strengthens the transition by lowering
$T_p$ and thereby increasing $\xi_p$.
To quantify the degree of supercooling, we adopt the percolation-based
measure~\cite{Athron:2022mmm}
\begin{equation}
\zeta_{\rm SC}
\equiv
1-\frac{T_p}{T_c}.
\end{equation}

Finally, we briefly comment on the bubble wall velocity $v_{\rm w}$,
which directly affects percolation.
Percolation is conventionally reached when the false-vacuum fraction
$P_{\rm false}(t)$ decreases to approximately $71\%$.
Since $P_{\rm false}(t)=e^{-I(t)}$ and $I(t)\propto v_{\rm w}^3$ for a
constant bubble wall velocity at fixed nucleation rate, a larger $v_{\rm w}$
causes the bubbles to fill space more rapidly.
Percolation is therefore reached earlier, at a higher temperature,
which tends to reduce $\xi_p$.
In this work, we adopt the benchmark value $v_{\rm w}=0.6$, commonly
used for non-runaway transitions~\cite{Megevand:2009gh,Cho:2026pfw}.

\section{Scanning Methodology and RGE analysis}
\label{sec-scan-method}

We first identify the Type-I parameter points that support an SFOEWPT
while satisfying the theoretical and experimental constraints.
To this end, we perform a comprehensive random scan over the six
independent 2HDM parameters
\begin{equation}
\label{eq-model-parameters}
\big\{
\tb,\; s_{\beta-\alpha},\; m_{12}^2,\;
m_{\varphi^0},\; \ma,\; \mch
\big\},
\end{equation}
where $\varphi^0$ denotes the BSM CP-even Higgs boson,
i.e., $\varphi^0=H$ in the NS and $\varphi^0=h$ in the IS.
For the NS, we adopt the following scan ranges:
\begin{equation}
\label{eq-scan-range-common}
\begin{alignedat}{2}
\ma &\in [15,2000]~\text{GeV}, &\qquad
\mch &\in [80,2000]~\text{GeV}, \\
\tb &\in [1,50], &
m_{12}^2 &\in [0,2\times10^6]~\text{GeV}^2 , \\
\mhh &\in [130,2000]~\text{GeV}, &
\qquad |s_{\beta-\alpha}| &\in [0.9,1],
\end{alignedat}
\end{equation}
and for the IS,
\begin{equation}
\label{eq-scan-range-scenario}
\begin{alignedat}{3}
\ma &\in [30,700]~\text{GeV}, &\qquad
\mch &\in [80,700]~\text{GeV}, \\
\tb &\in [1,50], &
m_{12}^2 &\in [0,2\times10^4]~\text{GeV}^2 , \\
\mh &\in [30,120]~\text{GeV}, &
\qquad |s_{\beta-\alpha}| &\in [0,0.5].
\end{alignedat}
\end{equation}

Throughout this work, we distinguish physical, FOEWPT, and SFOEWPT
parameter points.
A physical parameter point is required to satisfy both theoretical and
experimental constraints.
For theoretical consistency, we use the public package
\package{2HDMC}~\cite{Eriksson:2009ws} to impose vacuum
stability~\cite{Ivanov:2008cxa,Barroso:2012mj,Barroso:2013awa},
boundedness from below of the scalar potential~\cite{Ivanov:2006yq},
tree-level unitarity in scalar--scalar
scattering~\cite{Branco:2011iw,Arhrib:2000is}, and perturbativity of the
Higgs quartic couplings~\cite{Chang:2015goa}.
Experimentally, we impose the electroweak oblique-parameter constraints
$(S,T,U)$~\cite{Peskin:1991sw,He:2001tp,Grimus:2008nb,
ParticleDataGroup:2024cfk}, flavor constraints, in particular those from
$B_d\to\mu\mu$ in Type-I~\cite{Haller:2018nnx,Arbey:2017gmh,
Sanyal:2019xcp,Misiak:2017bgg}, precision measurements of the
$125~\text{GeV}$ SM-like Higgs signal strengths, and direct searches for
additional Higgs bosons at LEP, the Tevatron, and the LHC.
These constraints are implemented with \package{ScannerS}
version~2~\cite{Muhlleitner:2020wwk}, interfaced with
\package{HiggsTools}-v1.2~\cite{Bahl:2022igd}.

The thermal histories of the physical parameter points are evaluated
with \package{BSMPT} version 3.1.8~\cite{Basler:2018cwe,Basler:2020nrq,Basler:2024aaf}.
FOEWPT parameter points are those that undergo a first-order electroweak
phase transition irrespective of its strength, while SFOEWPT points
additionally satisfy $\xi_p>1$.
To ensure completion of the EWPT, we require all relevant
\package{BSMPT} status outputs
(\texttt{status\_nlo\_stability},
\texttt{status\_ewsr},
\texttt{status\_tracing},
\texttt{status\_coex\_pairs},
\texttt{status\_crit\_i},
\texttt{status\_bounce\_sol\_i},
\texttt{status\_nucl\_i},
\texttt{status\_perc\_i},
\texttt{status\_compl\_i}, and
\texttt{status\_gw\_i})
to return \texttt{success}.
In addition, we require the zero-temperature VEV to reproduce the
physical tree-level input,
\begin{equation}
\big|v-v^{\rm tree}\big|_{T=0}<1~\text{GeV}.
\end{equation}

Having identified the SFOEWPT parameter points, we next determine the
cutoff scale, defined as the highest energy scale up to which the 2HDM
remains theoretically consistent.
For this purpose, we monitor three conditions along the
renormalization-group flow---perturbativity, tree-level unitarity, and
vacuum stability~\cite{Machacek:1983tz,Machacek:1983fi,Machacek:1984zw,Haber:1993an,Luo:2002ti,Grimus:2004yh,Chakrabarty:2014aya,Das:2015mwa,Chakrabarty:2016smc,Kang:2022mdy}---and define the corresponding cutoff
scales $\Lambda_{\rm c}^{\rm pert}$, $\Lambda_{\rm c}^{\rm unit}$, and
$\Lambda_{\rm c}^{\rm stab}$.
The perturbativity cutoff $\Lambda_{\rm c}^{\rm pert}$ is the scale at
which the condition $|\lambda_i|<4\pi$ $(i=1,\ldots,5)$ is first
violated~\cite{Biekotter:2022kgf}.
The unitarity cutoff $\Lambda_{\rm c}^{\rm unit}$ is the scale at which
any eigenvalue of the tree-level scalar--scalar scattering matrices
exceeds $8\pi$, while the vacuum-stability cutoff
$\Lambda_{\rm c}^{\rm stab}$ is the scale at which the scalar potential
ceases to be bounded from below.
We define the overall cutoff scale as
\beq
\Lambda_{\rm c}
=
\min\left(
\Lambda_{\rm c}^{\rm pert},
\Lambda_{\rm c}^{\rm unit},
\Lambda_{\rm c}^{\rm stab}
\right).
\eeq
To determine these cutoff scales, we evolve the 2HDM parameters at the
two-loop level using the public code
\package{2HDME}~\cite{Oredsson:2018yho,Oredsson:2018vio}.

Finally, we define two benchmark requirements on the cutoff scale.
We adopt $\Lambda_{\rm c}>1\tev$ as a minimal UV-validity benchmark,
motivated by the absence of BSM signals in current LHC searches probing
the TeV scale~\cite{ATLAS:2024fdw,ATLAS:2024itc,ATLAS:2024vxm,CMS:2024yiy}.
We further consider the stronger requirement $\Lambda_{\rm c}>10\tev$,
which provides a clear separation between the electroweak and cutoff
scales.
At $\Lambda_{\rm c}=10\tev$,
$(v/\Lambda_{\rm c})^2\simeq6\times10^{-4}$, so higher-dimensional
effects are expected to be strongly suppressed, barring anomalously
large Wilson coefficients.
This hierarchy therefore reduces the expected impact of
higher-dimensional operators on the low-energy 2HDM
description~\cite{Contino:2016jqw,Kanemura:1999xf}.
The $10\tev$ scale is also relevant to the energy frontier envisioned
for future facilities such as a multi-TeV muon collider and the
FCC-hh~\cite{AlAli:2021let,Buttazzo:2020uzc}.

\section{Results}
\label{sec-results}

We first present an overview of the SFOEWPT samples, including their
thermal-history composition and survival under the cutoff-scale requirements.
Compared with our previous studies, the NS (IS) sample is increased from
$10^6$ ($2.36\times10^6$) physical parameter
points~\cite{Cho:2026pfw,Lee:2025hgb} to $5\times10^6$ points for each
scenario, improving the statistics particularly for rare multistep
transitions. The thermal histories of these samples are analyzed with
\package{BSMPT} and classified into FOEWPT and SFOEWPT parameter points.
While our previous IS analysis employed $v_{\rm w}=0.95$ and the
Arnold--Espinosa resummation scheme, here we adopt the common setup of
$v_{\rm w}=0.6$ and the Parwani scheme for both scenarios, enabling a
direct comparison between the NS and IS.

\begin{table}[t]
\centering
\setlength{\tabcolsep}{8pt}
\renewcommand{\arraystretch}{1.15}
\begin{tabular}{|c||r|c|r|r|r|}
\hline
Scenario
& $N_{\rm phy}$
& Transition
& $N_{\rm SFO}$
& $N_{\rm SFO}(\Lambda_{\rm c}>1\tev)$
& $N_{\rm SFO}(\Lambda_{\rm c}>10\tev)$ \\ \hline
\multirow{2}{*}{NS}
& \multirow{2}{*}{$5\times10^6$}
& one-step
& 419,118
& 28,415 (6.78\%)
& 520 (0.12\%) \\ \cline{3-6}
& & two-step
& 117
& 22 (18.8\%)
& 6 (5.1\%) \\ \hline
\multirow{3}{*}{IS}
& \multirow{3}{*}{$5\times10^6$}
& one-step
& 246,759
& 216,741 (87.84\%)
& 61,323 (24.85\%) \\ \cline{3-6}
& & two-step
& 1,718
& 1,715 (99.8\%)
& 1,280 (74.5\%) \\ \cline{3-6}
& & three-step
& 23
& 23 (100\%)
& 22 (95.7\%) \\ \hline
\end{tabular}
\caption{%
Numbers of SFOEWPT parameter points with one-, two-, and three-step
thermal histories in the Normal Scenario (NS) and Inverted Scenario (IS),
together with the numbers satisfying $\Lambda_{\rm c}>1\tev$ and
$\Lambda_{\rm c}>10\tev$.
The percentages in parentheses denote the surviving fractions relative to
the corresponding $N_{\rm SFO}$ sample.
}
\label{tab:sfoewpt-summary}
\end{table}

\autoref{tab:sfoewpt-summary} presents the numbers of SFOEWPT parameter
points with one-, two-, and three-step thermal histories in the NS and IS,
both before and after imposing the cutoff-scale requirements.
Out of the $5\times10^6$ physical parameter points generated for each
scenario, $419{,}235$ points in the NS and $248{,}500$ points in the IS
realize an SFOEWPT, corresponding to
$N_{\rm SFO}/N_{\rm phy}=8.385\%$ and $4.970\%$, respectively.
Thus, within our scan setup, SFOEWPTs occur more frequently in the NS.

For multistep thermal histories, we emphasize that $N_{\rm SFO}$ counts
parameter points rather than individual transition stages.
Specifically, for an $n$-step thermal history, a parameter point is classified
as an SFOEWPT point if at least one of the $n$ successive transition stages
satisfies $\xi_{p,i}>1$ $(i=1,\ldots,n)$.
If more than one stage satisfies this condition, the parameter point is
counted only once in $N_{\rm SFO}$.

With this definition, the SFOEWPT samples are overwhelmingly dominated by
one-step transitions.
In the NS, $99.972\%$ of the SFOEWPT points undergo a one-step transition,
while only $0.028\%$ proceed through two steps and no three-step SFOEWPT
is found.
In the IS, the corresponding one-, two-, and three-step fractions are
$99.300\%$, $0.691\%$, and $0.009\%$, respectively.
Multistep SFOEWPTs therefore occur more frequently in the IS than in the NS.
Although they remain rare even in the IS, the $1{,}718$ two-step SFOEWPT
points provide a sufficiently large sample for a dedicated study of their
characteristics.
A further interesting feature is that the largest $\xi_p$ is typically
attained in the final stage of a multistep transition.
This occurs for all two-step SFOEWPT points in the NS, for $97.4\%$ of the
two-step points in the IS, and for all three-step points in the IS.

Before imposing the cutoff-scale requirements, we examine which theoretical
condition predominantly determines $\Lambda_{\rm c}$.
Among the FOEWPT parameter points in the NS, tree-level unitarity
overwhelmingly sets the cutoff, accounting for $99.85\%$ of the points,
while vacuum stability accounts for $0.15\%$ and none is limited by
perturbativity.\footnote{In a restricted Type-II 2HDM parameter region
featuring SFOEWPTs, Ref.~\cite{Biekotter:2022kgf} found the perturbativity
cutoff to lie at $\Lambda_{\rm c}^{\rm pert}\sim1$--$2\tev$.}
In the IS, tree-level unitarity, vacuum stability, and perturbativity
determine the cutoff for $98.86\%$, $0.38\%$, and $0.77\%$ of the points,
respectively.

We next examine how the cutoff-scale requirements affect
the survival of the SFOEWPT parameter points.
In the NS, only $6.78\%$ of the one-step SFOEWPT parameter points survive
the minimal requirement $\Lambda_{\rm c}>1\tev$, and this fraction drops to
merely $0.12\%$ for $\Lambda_{\rm c}>10\tev$.
For the two-step SFOEWPTs, however, the corresponding survival fractions
increase to $18.8\%$ and $5.1\%$, respectively.
Thus, despite their rarity in the NS, two-step SFOEWPTs exhibit substantially
greater UV viability than their one-step counterparts.

This pattern becomes even more pronounced in the IS.
The one-step SFOEWPTs already exhibit large survival fractions under the
cutoff-scale requirements: $87.84\%$ satisfy $\Lambda_{\rm c}>1\tev$ and
$24.85\%$ satisfy $\Lambda_{\rm c}>10\tev$, both substantially higher
than the corresponding fractions in the NS.
The survival fractions are even larger for multistep transitions.
Specifically, $99.8\%$ ($74.5\%$) of the two-step points satisfy
$\Lambda_{\rm c}>1\tev$ ($10\tev$), while all $23$ three-step points
survive the $1\tev$ requirement and $22$ remain viable beyond $10\tev$.
Hence, although the NS contains a larger overall SFOEWPT population before
the cutoff-scale requirements are imposed, the IS retains far more
UV-valid SFOEWPT points, with multistep transitions exhibiting
particularly high survival fractions.

\begin{figure}[t]
\centering
\includegraphics[width=\textwidth]{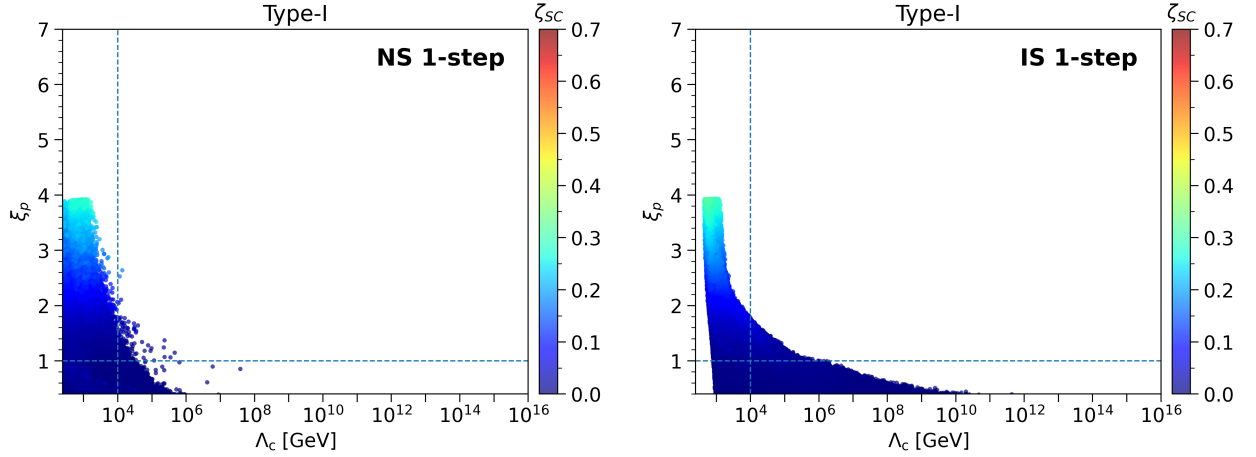}
\caption{%
Distributions of the one-step FOEWPT parameter points in the
$(\Lambda_{\rm c},\xi_p)$ plane for the Type-I 2HDM in the NS (left) and
IS (right).
The Parwani resummation scheme and $v_{\rm w}=0.6$ are adopted.
The color scale denotes the supercooling measure
$\zeta_{\rm SC}\equiv1-T_p/T_c$.
The dashed lines mark $\Lambda_{\rm c}=10\tev$ (vertical) and the SFOEWPT
condition $\xi_p=1$ (horizontal).
}
\label{fig-xip-Lm-onestep-NSIS}
\end{figure}

We now turn to the relation between the EWPT strength and the cutoff scale.
\autoref{fig-xip-Lm-onestep-NSIS} presents all one-step FOEWPT parameter
points in the $(\Lambda_{\rm c},\xi_p)$ plane, rather than restricting the
sample to $\xi_p>1$, in order to display the full correlation between
$\xi_p$ and $\Lambda_{\rm c}$.
The left and right panels correspond to the NS and IS, respectively.
The color scale represents the supercooling measure $\zeta_{\rm SC}$.
The vertical and horizontal dashed lines indicate
$\Lambda_{\rm c}=10\tev$ and the SFOEWPT threshold $\xi_p=1$,
respectively.

The most striking feature is the pronounced anticorrelation exhibited by
the upper envelope of both distributions.
Imposing a larger minimum cutoff scale progressively lowers the maximal
allowed transition strength.
In particular, requiring $\Lambda_{\rm c}>10\tev$ limits the one-step
transition to approximately $\xi_p\lesssim2.7$ in the NS and
$\xi_p\lesssim1.8$ in the IS.
Conversely, within our scan, requiring $\xi_p>1$ restricts the cutoff scale
to $\Lambda_{\rm c}\lesssim4.6\times10^5~\text{GeV}$ in the NS and
$\Lambda_{\rm c}\lesssim9.6\times10^5~\text{GeV}$ in the IS.
Thus, a one-step SFOEWPT in the Type-I 2HDM substantially limits the energy
range over which the 2HDM remains theoretically consistent.

The color distributions in \autoref{fig-xip-Lm-onestep-NSIS} further reveal
two correlations involving supercooling.
First, larger $\zeta_{\rm SC}$ is associated with larger $\xi_p$.
This positive correlation was demonstrated for the NS in our previous
study~\cite{Cho:2026pfw}, and the present results show that it persists
in the IS.
Second, and more notably, stronger supercooling is preferentially associated
with a lower cutoff scale.
For $\Lambda_{\rm c}>1\tev$, the maximal supercooling is approximately
$\zeta_{\rm SC}\simeq0.3$, whereas for $\Lambda_{\rm c}>10\tev$ it falls
below approximately $0.1$.
Increasing the required UV-validity scale therefore not only limits the
transition strength but also strongly disfavors substantial supercooling.

\begin{figure}[t]
\centering
\includegraphics[width=\textwidth]{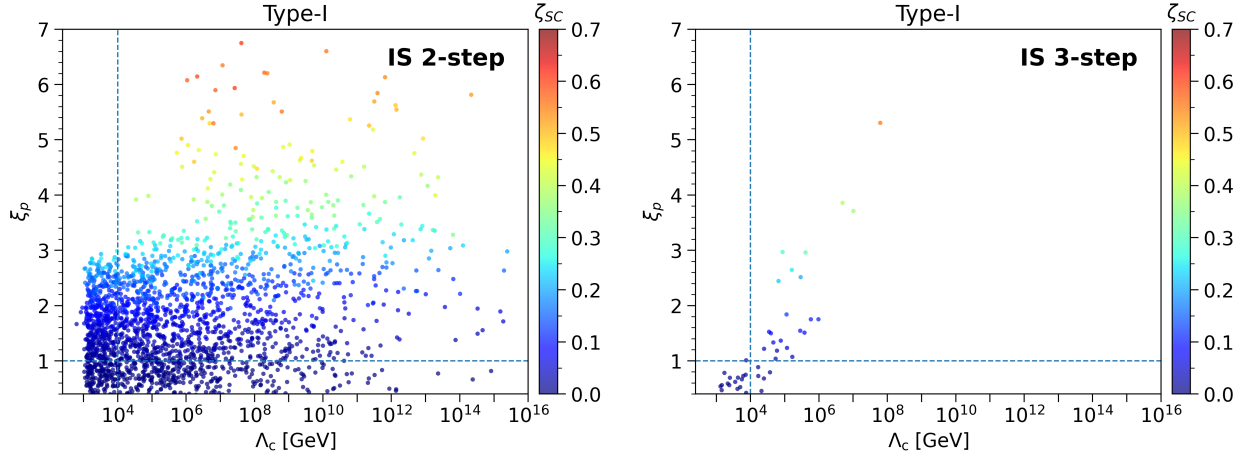}
\caption{%
Distributions of the FOEWPT parameter points in the
$(\Lambda_{\rm c},\xi_p)$ plane for two-step (left) and three-step
(right) thermal histories in the IS for the Type-I 2HDM.
The color scale denotes $\zeta_{\rm SC}\equiv1-T_p/T_c$, and the dashed
lines are as in \autoref{fig-xip-Lm-onestep-NSIS}.
}
\label{fig-xip-Lm-23step-IS}
\end{figure}

Multistep transitions exhibit qualitatively different behavior.
\autoref{fig-xip-Lm-23step-IS} shows the corresponding
$(\Lambda_{\rm c},\xi_p)$ distributions for two-step (left) and three-step
(right) thermal histories in the IS.
Most notably, two-step SFOEWPTs can remain theoretically consistent up to
scales of order $2.4\times10^{15}~\text{GeV}$, comparable to typical GUT scales.
Three-step SFOEWPTs reach lower maximal cutoff scales of approximately
$6.2\times10^7~\text{GeV}$.
The Type-I 2HDM, when realizing a two-step SFOEWPT, can therefore remain
theoretically consistent up to very high UV scales without requiring an
intermediate-scale UV completion from these consistency conditions alone.

Equally importantly, the pronounced anticorrelation between
$\Lambda_{\rm c}$ and $\xi_p$ in the upper envelope of the one-step
distributions is not observed for multistep transitions.
Unlike in the one-step case, increasing $\Lambda_{\rm c}$ does not
systematically lower the maximal allowed $\xi_p$, indicating that a
strong multistep transition need not be associated with a low cutoff scale.
Moreover, substantial supercooling remains possible, with
$\zeta_{\rm SC}$ reaching approximately $0.7$ for the two-step sample.

The three-step sample is much smaller, and its quantitative features
should therefore be interpreted with caution.
Nevertheless, the surviving points reach cutoff scales of approximately
$6.2\times10^7\gev$, transition strengths up to $\xi_p\sim5.3$, and
supercooling as large as $\zeta_{\rm SC}\simeq0.554$.
Within this limited sample, no clear $\Lambda_{\rm c}$--$\xi_p$
anticorrelation is apparent either, suggesting that the tension between
EWPT strength and UV validity found for one-step transitions is not a
generic feature of multistep thermal histories.

\begin{figure}[t]
\centering
\includegraphics[width=\textwidth]{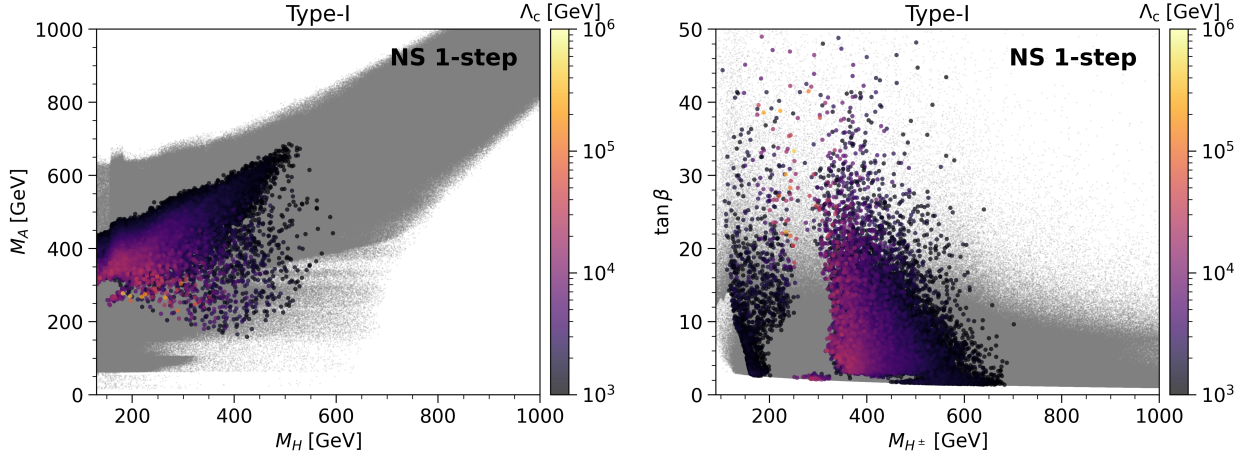}
\caption{%
Distributions of the UV-viable one-step SFOEWPT parameter points in the
$(M_H,M_A)$ plane (left) and the $(M_{H^\pm},\tan\beta)$ plane (right)
for the NS of the Type-I 2HDM with the Parwani resummation scheme.
Colored points satisfy the minimal UV-validity requirement
$\Lambda_{\rm c}>1\tev$, with the color scale indicating
$\Lambda_{\rm c}$, while the underlying gray points denote the physical
parameter points.
}
\label{fig-NS-1step}
\end{figure}

We next examine the characteristics of SFOEWPT parameter points satisfying
the minimal UV-validity requirement $\Lambda_{\rm c}>1\tev$, focusing in
particular on the relation between the cutoff scale and the BSM Higgs mass
scales.
\autoref{fig-NS-1step} shows the $(M_H,M_A)$ and
$(M_{H^\pm},\tb)$ distributions for one-step SFOEWPTs in the NS.
In the $(M_H,M_A)$ plane, the UV-viable points populate a correlated region
at moderate scalar masses, with approximately
$M_A \in [150,680]~\text{GeV}$ and $M_H\in [130,600]~\text{GeV}$.
Notably, the cutoff scale tends to increase toward the lower-mass part of
this region, indicating that greater UV validity does not favor a heavier
BSM Higgs spectrum.

In the $(M_{H^\pm},\tb)$ plane, the $\Lambda_{\rm c}>1\tev$ requirement
has a pronounced impact on the charged-Higgs mass, restricting it to
approximately $M_{H^\pm}\lesssim700~\text{GeV}$, whereas $\tb$ remains
comparatively weakly constrained, with viable points extending up to
$\tb\simeq50$.
The surviving points exhibit three characteristic structures: a dominant
region around $M_{H^\pm}\simeq300$--$700~\text{GeV}$ spanning a broad range of $\tb$; a
nearly vertical branch around $M_{H^\pm}\simeq 100$--$250~\text{GeV}$ extending
toward relatively large $\tb$; and a small isolated island near
$M_{H^\pm}\sim300~\text{GeV}$ and $\tb\sim3$.
The cutoff scale tends to increase toward
$M_{H^\pm}\sim300$--$450~\text{GeV}$, indicating that greater UV validity
preferentially selects an intermediate charged-Higgs mass range.

\begin{figure}[t]
\centering
\includegraphics[width=\textwidth]{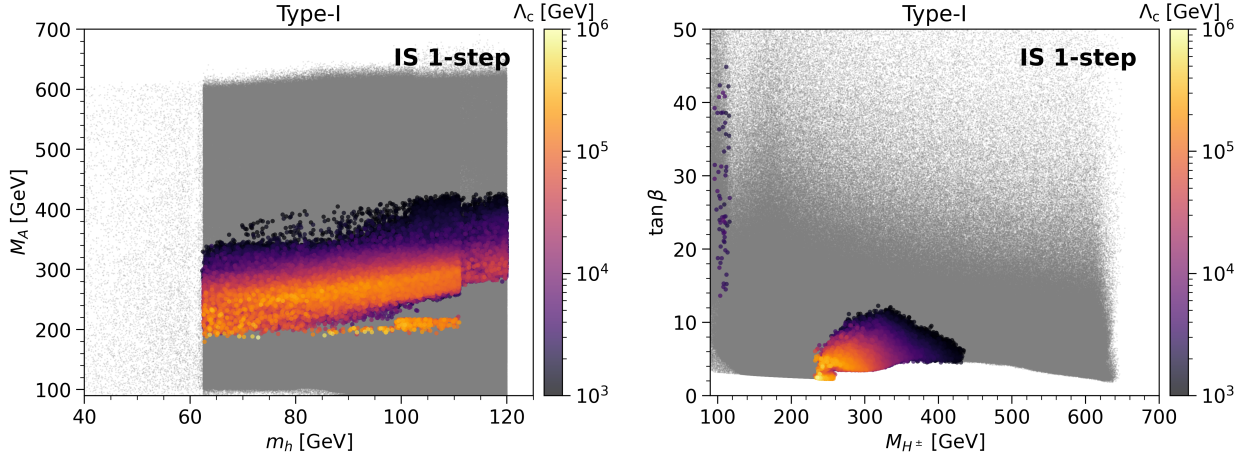}
\caption{%
Same as \autoref{fig-NS-1step} for the one-step SFOEWPT in the IS.
}
\label{fig-IS-1step}
\end{figure}

The corresponding one-step distributions in the IS are shown in
\autoref{fig-IS-1step}.
In the $(m_h,M_A)$ plane, requiring $\Lambda_{\rm c}>1\tev$ restricts the
pseudoscalar mass to approximately $M_A\in[190,425]\gev$, while leaving
$m_h$ essentially unconstrained beyond the existing experimental limits.
The depletion of viable points for $m_h\lesssim m_{h_{125}}/2$ is mainly
due to the stringent constraints on the exotic decay
$h_{125}\to hh$.
As the minimum required cutoff scale is increased, the viable
pseudoscalar-mass range progressively narrows and shifts toward lighter
$M_A$.

The $(M_{H^\pm},\tb)$ distribution is more restricted than in the NS.
The dominant region is confined to
$M_{H^\pm}\simeq230$--$440\gev$ and $\tb\lesssim12$, while the low-mass
branch is reduced to a sparse, nearly vertical strip around
$M_{H^\pm}\sim100\gev$ with $\tb\gtrsim14$.
The isolated low-$\tb$ island present in the NS is absent in the IS.
The cutoff scale tends to increase toward
$M_{H^\pm}\sim250\gev$ and low $\tb$, further localizing the viable
one-step SFOEWPT parameter space in a region where improved flavor
measurements may provide complementary probes.

\begin{figure}[t]
\centering
\includegraphics[width=\textwidth]{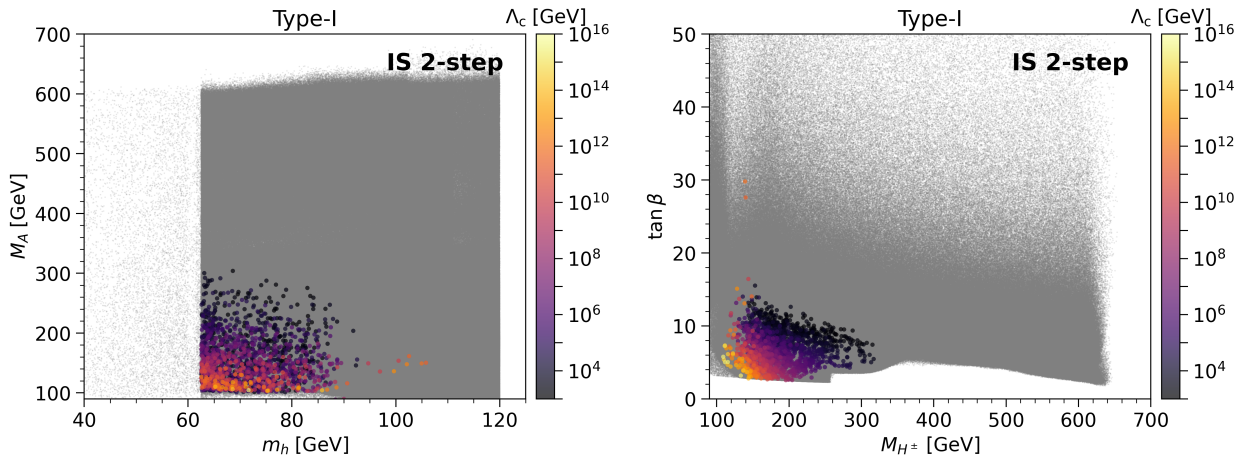}
\caption{%
Same as \autoref{fig-NS-1step} for the two-step SFOEWPT in the IS.
}
\label{fig-IS-2step}
\end{figure}

Finally, \autoref{fig-IS-2step} presents the $(m_h,M_A)$ and
$(M_{H^\pm},\tb)$ distributions for two-step SFOEWPTs in the IS.
Both distributions differ markedly from their one-step counterparts.
In the $(m_h,M_A)$ plane, the minimal requirement
$\Lambda_{\rm c}>1\tev$ selects substantially lighter scalar masses.
In particular, $M_A$ extends down to approximately $100\gev$, while its
upper bound lies near $300\gev$.
Unlike in the one-step case, the light CP-even Higgs mass $m_h$ is also
strongly affected by the cutoff-scale requirement.
Most points are concentrated in
$m_h\simeq62.5$--$90\gev$, with only a small fraction extending up to
approximately $106\gev$.

The $(M_{H^\pm},\tb)$ distribution also differs significantly from its
one-step counterpart.
The main compact region persists but shifts toward lower charged-Higgs
masses, with $M_{H^\pm}\lesssim320\gev$, while the sparse vertical branch
around $M_{H^\pm}\sim100\gev$ disappears.
Almost all viable two-step points satisfy $\tb\lesssim16$.

As the minimum required cutoff scale is increased, both distributions
shift toward lighter BSM Higgs masses and moderate $\tb$.
For example, imposing $\Lambda_{\rm c}>10^{13}\gev$ restricts the
surviving points to
$M_A\in[64.6,117.7]\gev$,
$m_h\in[64.4,85.8]\gev$,
$M_{H^\pm}\in[109.1,147.2]\gev$,
and $\tb\in[2.8,7.3]$.
In the Type-I 2HDM, this moderate-$\tb$ region also entails weaker
Yukawa suppression than at large $\tb$.
Together with the light BSM Higgs spectrum, this makes the
very-high-cutoff two-step SFOEWPT region a promising target for future
high-energy collider searches.

\section{Conclusions}
\label{sec-conclusions}

We have investigated the UV viability of SFOEWPTs in the Type-I 2HDM by
combining an extensive parameter scan with a two-loop RGE analysis of the cutoff
scale $\Lambda_{\rm c}$.
For each of the Normal and Inverted Scenarios, we scanned
$5\times10^6$ physical parameter points, identified SFOEWPT points with
one-, two-, and three-step thermal histories, and determined the cutoff scale at
which perturbativity, tree-level unitarity, or vacuum stability first
fails.
Tree-level unitarity determines the cutoff for the overwhelming majority
of points in both scenarios.

Our main result is the sharp contrast between one-step and multistep
thermal histories.
For one-step transitions, the EWPT strength and UV validity are in direct
tension: requiring $\Lambda_{\rm c}>10\tev$ limits the transition strength
to $\xi_p\lesssim2.7$ in the NS and $\xi_p\lesssim1.8$ in the IS, while
requiring $\xi_p>1$ restricts the cutoff to
$\Lambda_{\rm c}\lesssim\mathcal{O}(10^6)~\text{GeV}$ in both scenarios.
Thus, a one-step SFOEWPT substantially limits the UV range over which the
Type-I 2HDM can remain theoretically consistent.
Multistep transitions behave qualitatively differently.
Although rare---accounting for $0.028\%$ of the SFOEWPT points in the NS
and $0.700\%$ in the IS---they do not exhibit the
$\Lambda_{\rm c}$--$\xi_p$ anticorrelation found for one-step transitions.
In particular, two-step SFOEWPTs in the IS can remain theoretically
consistent up to $\Lambda_{\rm c}\sim\mathcal{O}(10^{15})~\text{GeV}$, while
allowing $\xi_p\simeq7$ and substantial supercooling.
The tension between EWPT strength and UV validity is therefore characteristic
of one-step thermal histories rather than a generic feature of SFOEWPTs.

Imposing cutoff-scale requirements also sharpens the phenomenologically
viable parameter space.
In the NS, requiring $\Lambda_{\rm c}>1\tev$ yields approximate upper
bounds of $M_H\lesssim600\gev$, $M_A\lesssim680\gev$, and
$M_{H^\pm}\lesssim700\gev$.
Increasing the minimum required cutoff scale further favors lower
neutral-scalar masses and an intermediate charged-Higgs mass range.
This selection is particularly pronounced for two-step SFOEWPTs in the IS.
For example, requiring $\Lambda_{\rm c}>10^{13}\gev$ restricts the
surviving points to
$M_A\in[64.6,117.7]\gev$,
$m_h\in[64.4,85.8]\gev$,
$M_{H^\pm}\in[109.1,147.2]\gev$,
and $\tb\in[2.8,7.3]$.
This very-high-cutoff two-step region is therefore characterized by a
light BSM Higgs spectrum and moderate $\tb$, making it a promising target
for the HL-LHC and future high-energy colliders such as a multi-TeV muon
collider and the FCC-hh.

\acknowledgments
The work of JC and DK is supported by National Institute for Mathematical Sciences (NIMS) grant funded by the Korea government (MSIT) (No.~B26810000).
The work of JK is supported by a KIAS Individual Grant (PG099202) and by the Center for Advanced Computation, both at the Korea Institute for Advanced Study.
The work of JS is supported by the National Research Foundation of Korea, Grant
No.~RS-2026-25588929.

\bibliographystyle{JHEPMod}
\bibliography{RGE-2HDM-SFOEWPT}

\end{document}